\documentclass{IEEEtran4PSCC}

\ifCLASSINFOpdf
   \usepackage[pdftex]{graphicx}
\else
   \usepackage[dvips]{graphicx}
\fi

\usepackage[cmex10]{amsmath}

\makeatletter
\let\old@ps@headings\ps@headings
\let\old@ps@IEEEtitlepagestyle\ps@IEEEtitlepagestyle
\def\psccfooter#1{%
    \def\ps@headings{%
        \old@ps@headings%
        \def\@oddfoot{\strut\hfill#1\hfill\strut}%
        \def\@evenfoot{\strut\hfill#1\hfill\strut}%
    }%
    \def\ps@IEEEtitlepagestyle{%
        \old@ps@IEEEtitlepagestyle%
        \def\@oddfoot{\strut\hfill#1\hfill\strut}%
        \def\@evenfoot{\strut\hfill#1\hfill\strut}%
    }%
    \ps@headings%
}
\makeatother

\usepackage{mathrsfs}%
\usepackage{amsmath,amsfonts}
\usepackage{algorithm}
\usepackage{amsthm}
\usepackage{amssymb}
 \usepackage{algpseudocode}
\usepackage{array}
\usepackage{textcomp}
\usepackage{stfloats}
\usepackage{subfig}
\usepackage{float}
\usepackage{url}
\usepackage{verbatim}
\usepackage{graphicx}
\usepackage{cite}
\usepackage{eurosym}

\newtheorem{definition}{Definition}
\usepackage{multirow}
\usepackage{hyperref}
\usepackage{enumitem}
\usepackage{graphicx}
\usepackage{nomencl}
\hypersetup{
	colorlinks,
	citecolor=blue,
	filecolor=blue,
	linkcolor=blue,
	urlcolor=blue}
\usepackage[compact]{titlesec}         % you need this package
\titlespacing{\section}{0pt}{0pt}{0pt} % this reduces space between (sub)sections to 0pt, for example
\AtBeginDocument{%                     % this will 
  \setlength\abovedisplayskip{0pt}
  \setlength\belowdisplayskip{0pt}}
  
\usepackage[table]{xcolor} 
\makenomenclature

\usepackage{tikz}
\usetikzlibrary{matrix}
\usepackage{blindtext}
\usepackage{multicol}
\usepackage{adjustbox}
\usepackage{mathtools}
\usepackage{booktabs}
\usepackage{graphicx}
\usepackage{caption}
\usepackage{mwe}
\usepackage{nomencl}
\makenomenclature

\begin{document}

\setlength{\abovedisplayskip}{5pt}
\setlength{\belowdisplayskip}{5pt}

%\title{Learning Price in Day-ahead Electricity Market}

\title{Learning  to Bid in Forward Electricity Markets \\ Using a No-Regret Algorithm}

\author{
\IEEEauthorblockN{Arega Getaneh Abate$^{a}$, Dorsa Majdi{$^b$}, Jalal Kazempour$^{a}$, and Maryam Kamgarpour$^{c}$} 
\IEEEauthorblockA{$^{a}$ Department of Wind and Energy Systems, Technical University of Denmark, 
 Kgs. Lyngby, Denmark}
 
 \IEEEauthorblockA{$^{b}$ Sharif University of Technology, Tehran, Iran}
 
\IEEEauthorblockA{$^{c}$ Sycamore lab, École Polytechnique Fédérale de Lausanne, Lausanne, Switzerland}

\IEEEauthorblockA{ \{ageab, jalal\}@dtu.dk, maryam.kamgarpour@epfl.ch, dorsa.majdi@gmail.com }

\vspace{-7mm}
}
% \thanks{\IEEEauthorrefmark{a}Corresponding author.}
\maketitle

\begin{abstract}
It is a common practice in the current literature of electricity markets to use game-theoretic approaches for strategic price bidding. However, they generally rely on the assumption that the strategic bidders have prior knowledge of rival bids, either perfectly or with some uncertainty. This is not necessarily a realistic assumption. This paper takes a different approach by relaxing such an assumption and exploits a no-regret learning algorithm for repeated games. In particular, by using the \emph{a posteriori} information about rivals' bids, a learner can implement a no-regret algorithm to optimize her/his decision making. Given this information, we utilize a multiplicative weight-update algorithm, adapting bidding strategies over multiple rounds of an auction to minimize her/his regret. 
Our numerical results show that when the proposed learning approach is used the social cost and the market-clearing prices can be higher than those corresponding to the classical game-theoretic approaches. The takeaway for market regulators is that electricity markets might be exposed to greater market power of suppliers than what classical analysis shows. 
\end{abstract}


\vspace{4mm}
\section{Introduction}
\vspace{1mm}

\subsection{Background and Motivation}
\vspace{-1mm}
Strategic bidding is a key task in the daily operation of electricity suppliers, so-called \textit{bidders}. There is a vast literature, mostly built upon game-theoretic approaches such as bi-level programming, allowing bidders to derive their strategic bids in terms of price or quantity or both \cite{diag,hobbs,martin}. One critical assumption in such works is that the bidder has knowledge of the decision making process of rival bidders, market price formation, and thereby the market outcomes for every bidder. While some works model the uncertainty on rival bids \cite{ruiz2009pool}, they still assume that the bidder perfectly knows how the market is being cleared.  

These assumptions do not necessarily hold in real-world electricity auctions for several reasons, such as (\textit{i}) bidding decisions are private to bidders, (\textit{ii}) bidders typically only have access to public post-auction information such as aggregate demand and supply curves and market-clearing prices, and (\textit{iii}) electricity auctions often involve multiple rounds and different set of bidders in each round. Therefore, formulating bidding problems with such unrealistic assumptions may eventually result in suboptimal decisions \cite{auer2002finite,wang2022earning} or outcomes that do not match the real world. 

This paper adopts an approach based on \textit{learning in repeated games} \cite{Nachbar}, and leverages no-regret learning from the online optimization literature \cite{orabona2019modern,shalev2012online} to address this problem. There has been a growing body of work that leverages online and reinforcement learning for bidding in electricity markets. In the following section, we highlight the related works.

\subsection{Literature Review}
In \cite{wang2022earning}, the authors investigate the sequential optimization of bidders to maximize payoffs using an adversarial multiarmed bandit-based real-time bidding scheme in an electricity market. For strategic bidding in electricity markets, a deep reinforcement learning method is developed in \cite{Yujian}, where physical non-convex operating characteristics of generators are accounted for. Considering unobservable information, \cite{du2021approximating} develops a multi-agent policy gradient algorithm aimed at approximating the Nash equilibrium among multiple strategic bidders in the day-ahead electricity market. The authors conclude that their algorithm enables all the market participants to find superior bidding strategies with increased profit gains. Reference \cite{karaca2020no} considers online learning in the general class of repeated electricity market auctions, and presents an approach to incorporate extra information available when a bid is not accepted, to estimate the utility function of a bidder.
A learning-based approach is developed in \cite{zhao2018electricity} using a multi-armed bandit algorithm for real-time pricing in a demand response program. Finally, \cite{baltaoglu2018algorithmic} utilizes online learning approaches to address the problem of optimal bidding for virtual trading in two-settlement electricity markets. 

To the best of our knowledge, none of the above work, and more broadly existing studies on learning in electricity markets, have considered the day-ahead electricity market auction in light of public post-market clearing data. In particular, in several current day-ahead markets, after each round of the auction, each bidder not only observes the outcomes of her/his submitted bids but also gets access to (\textit{i}) hourly market-clearing price, (\textit{ii}) hourly buy and sell volumes, and (\textit{iii}) hourly aggregate supply and demand curves. Such feedback information can be used by the bidder to assess potential outcomes of her/his alternative price bids\footnote{Please see the day-ahead aggregate supply and demand curves published by European market operators, such as EPEX \cite{EPEXspot} and Nord Pool \cite{Nordpool}.}. We refer to such an information setting as full information feedback. 

While advanced learning algorithms can be designed and tailored to a specific electricity market, our goal in this paper is to use an off-the-shelf online learning algorithm under full information feedback to understand the effect of online learning on social welfare and the learner's utility.

\subsection{Our Contributions}
\vspace{-1mm}

Our learning approach aims to serve as a benchmark to analyze how much market power bidders could potentially exert in a day-ahead market. To this end, we employ full information feedback and a simple market setting for which we can compute the Nash equilibrium. This enables us to compare social welfare and bidders' payoffs under the learning strategy with that corresponding to the Nash equilibrium strategy. The learning algorithm employed is the \textit{Hedge}  algorithm \cite{hedge} and enables the bidder over time, that is, over auction rounds, to effectively learn how to place a strategic price bid in a day-ahead electricity market {without} making any assumption on knowing the rivals' true costs or their next actions a priori. 

To compare the learning results with existing approaches in the literature, we consider each bidder solving a bi-level program, perfectly knowing the best response of rival bidders, leading to an equilibrium problem with equilibrium constraints (EPEC)\cite{Ehrenmann}. This problem can then be solved by a diagonalization method \cite{diag0,diag}. In our case study, we verify that the approach converges to a Nash equilibrium strategy and hence, serves as a benchmark for our learning approach. 

Our study illustrates the extent of social welfare suboptimality when an online learning approach is employed. Furthermore, it illustrates the cases for which a given bidder can improve her/his utilities by employing such a learning approach. Last, it also highlights the limitations of the no-regret criteria when used to infer bidders' utilities in an electricity market auction.

The rest of the paper is organized as follows.  Section \ref{perlim} provides preliminaries on repeated electricity market auctions. Section \ref{med} lays out the proposed no-regret algorithm. Section \ref{bench} formulates the benchmark model and its solution method. Section \ref{case} presents illustrative case studies.  Section \ref{conc} concludes the paper.

\vspace{3mm}
\section{Repeated Electricity Market Auction}\label{perlim}
We study a day-ahead electricity market and consider the electricity market auction as our stage game. The set of participants consists of the power generators referred to as \textit{Bidders} $\ell \in \mathcal{N}=\{1, \ldots,|\mathcal{N}|\}$ and an auctioneer or \textit{Market Operator} whose goal is to procure the total market demand, modeled by a price-inelastic demand $Q \in \mathbb{R}_{+}$, from the bidders at the minimum cost. We assume each bidder has a private \textit{true} cost function $C_{\ell}: \mathbb{X}_{\ell} \rightarrow \mathbb{R}_{+}$, $\mathbb{X}_{\ell} \subseteq \mathbb{R}_{+}$. We further assume that $0 \in \mathbb{X}_{\ell}$ and $C_{\ell}(0)=0$ and each bidder's cost function is quadratic, therefore is of the form $C_{\ell}(x) = \frac{1}{2}c_{\ell} x^2 + d_{\ell} x$ \cite{sessa2019no}. Each bidder $\ell$ has a finite strategy set $\mathcal{K}_{\ell}=\left\{1, \ldots,\left|\mathcal{K}_{\ell}\right|\right\}$ that consists of her/his true cost function and bid functions of the form $b_{\ell}^k,$ where $b_{\ell}^k(x) =\frac{1}{2}c^k_{\ell} x^2 + d^k_{\ell} x$. Without loss of generality, $1 \in \mathcal{K}_{\ell}$ corresponds to the true cost function.

Let $T\!\in\!\mathbb{N}$ be the time horizon length, i.e., the number of rounds the underlying auction is repeated. To illustrate the performance of the algorithm in a benchmark setting, we assume that for all rounds, the total demand $Q$, the true costs, and the strategy set $\mathcal{A}_{\ell}$ remain unchanged. Let $k_{\ell,t} \in \mathcal{K}_{\ell}$ denote the strategy of bidder $\ell$ at time $t \leq T$ which is only revealed to the market operator and ex-post published by the market operator anonymously as a part of the aggregate supply curve. Given the strategy profile $\mathcal{B}_t=\left\{b_{\ell}^{k_{\ell,t}}\right\}_{\ell \in \mathcal{N}}$, a mechanism defines an allocation rule $x_{\ell}^*(\mathcal{B}_t) \in \mathbb{X}_{\ell}$, and a payment rule $p_{\ell}(\mathcal{B}_t) \in \mathbb{R}$ for each bidder $\ell$. In many auctions, the allocation rule is determined by an optimization problem
\begingroup
\allowdisplaybreaks
\begin{subequations} \label{auction}
\begin{align}
J(\mathcal{B}_t)= & \min _{x \in \mathbb{X}_t} \sum_{\ell \in \mathcal{N}} b_{\ell}^{k_{\ell,t}}\left(x_{\ell}\right) \\
& \text { s.t. } \sum_{\ell \in \mathcal{N}} x_{\ell} \in \mathbb{S},
\end{align}
\end{subequations}
where $\mathbb{X}=\prod_{\ell \in \mathcal{N}} \mathbb{X}_{\ell} \subset \mathbb{R}_{+}^{|\mathcal{N}|}$ is the decision set for the optimization problem, and the set $\mathbb{S} \subset \mathbb{R}_{+}$ corresponds to the market constraints.

In an electricity market auction, these constraints in their simplest form may correspond to the capacity constraints of bidders and the total demand of the market. Hence, the market-clearing optimization problem solved by the market operator can be expressed as
\begin{subequations}\label{constraints}
\begin{align}
 \min _{x \in \mathbb{X}} \quad & \sum_{\ell \in \mathcal{N}} \frac{1}{2}c_{\ell}^{k_{\ell,t}}x_{\ell}^2 + d_{\ell}^{k_{\ell,t}}x_{\ell} \label{constraints1} \\
  \text { s.t. } \quad & \sum_{\ell} x_{\ell} = Q, \label{constraints2} \\
  & 0 \leq x_{\ell}\leq \Bar{x}_{\ell} \ \forall{\ell \in \mathcal{N}}.
  \end{align}
\end{subequations}

Let the optimal solution of \eqref{constraints} be denoted by $x^*(\mathcal{B}_t)$. Let $\lambda^*(\mathcal{B}_t)\!\in\!\mathbb{R}$ denote the Lagrange multiplier associated with \eqref{constraints2}. 
In the context of the electricity market auction, $\lambda^*(\mathcal{B}_t)$ is called the marginal price and is announced after each round $t$ to all bidders. The payment for each bidder $\ell$ is  $p_{\ell}(\mathcal{B}_t)=\lambda^*(\mathcal{B}_t)x_{\ell}^*$. The utility of bidder $\ell$ is linear in the payment received; $u_{\ell,t}=u_{\ell}(\mathcal{B}_t)=p_{\ell}(\mathcal{B}_t)-C_{\ell}\left(x_{\ell}^*\right)$. A bidder whose bid is not accepted i.e., $x_{\ell}^*=0$, is not paid, and $u_{\ell,t}=0$.

While Nash equilibrium offers a theoretical solution concept, real-world bidders are profit-maximizing entities operating within constraints of privacy and limited information. Consequently, it may not be practical to assume that they can compute or would be willing to select their Nash equilibrium strategy. Instead, a more {pragmatic} assumption is that bidders choose their strategies using some adaptive algorithm based on observed auction data. In the current practice of European day-ahead electricity markets, market operators publish ex-post hourly aggregate supply and demand curves (see, e.g., \cite{EPEXspot} and \cite{Nordpool}). This information could be leveraged to enable each bidder to retrospectively calculate her/his payoff for any chosen bid. The possibility of a posteriori evaluating the payoff function is referred to as \emph{full information feedback} in online learning,  in contrast to the so-called bandit feedback where the bidder  evaluates her/his payoff only for the submitted bid. 

\textbf{Remark:} The considered setup abstracts several realistic considerations, including time-varying demands,  different sets of possible bids and bidders in each round of the auction, different types of bid curves such as piece-wise constant, and merit order dispatch approaches. With our simplifications, we are able to benchmark the learning approach and provide a deeper understanding of the potential of exploiting posterior information feedback in the market under consideration.

\vspace{4mm}
\section{No-Regret Learning Approach}\label{med}
\vspace{1mm}
\subsection{Regret and connections to game}
From the perspective of a learning bidder $\ell$, the problem of learning in a repeated auction can be cast as an instance of online learning. In such a setting, the learner is faced with a sequence of time-varying utility functions and she/he aims to maximize her/his cumulative utility.  Note that in our auction setting, the time variations on the utility of bidder $\ell$ are induced due to the bidders changing strategies at each auction round. 

Recall $\mathcal{B}_t$ is the strategy profile chosen at auction round $t$ and let $\mathcal{B}_{-\ell,t}=$ $\left\{b_j^{k_{j,t}}\right\}_{j \in \mathcal{N} \backslash\{\ell\}}$.
The regret of bidder $\ell$ at time $T$ is defined as follows:
\begin{definition}[Regret] The \textit{regret} of  bidder $\ell$ at time $T$ is
\begin{align}
R_{\ell,T}=\max_{k \in \mathcal{K}_{\ell}} \ \sum_{t=1}^T u_{\ell}\left(\left\{\mathcal{B}_{-\ell,t}, b_{\ell}^k\right\}\right)-\sum_{t=1}^T u_{\ell,t}.
\end{align}
\end{definition}
Observe that after $T$ rounds, $R_{\ell, T}$ measures the difference between two quantities. The first term quantifies the maximum payoff bidder $\ell$ could have made had the bidder known the sequence of rival bids ahead of time, and had the bidder chosen the best fixed strategy in $\mathcal{K}_{\ell}$. The second term represents the cumulative utility derived from the strategy employed by bidder $\ell$ throughout the rounds. An algorithm for bidder $\ell$ is no-regret if $R_{\ell,T} / T \rightarrow 0$ as $T \rightarrow \infty$. 

Note that while no-regret algorithms can be defined for the case in which the sequence of received payoffs is arbitrary, they have additional connections to equilibria in a game setting. In particular, if every participant bids according to a no-regret algorithm, the empirical distribution of participants' bids converges to a coarse-correlated equilibrium, a relaxation of the Nash equilibria, of the one-shot game \cite{cesa2006prediction}.

\subsection{No-regret algorithm}
We consider the case in which, after each round of the auction, the bidder can accurately calculate her/his payoff had the bidder submitted alternative bids, namely, the full information feedback. In practice, this can be done based on the information published by the market operator after the day-ahead market clearing. This feedback allows the bidder to observe a vector of rewards. To exploit this information, we implement the \textit{Hedge} algorithm, modeling the bidder's reward function based on historical rounds of the auction. This algorithm enables the bidder to sequentially and adaptively learn how to maximize her/his \emph{cumulative} payoff by strategically selecting her/his bids across multiple auction rounds.

\begin{algorithm}[t]
\caption{Hedge algorithm for bidder $\ell$}\label{alg_cap}
\begin{algorithmic}[1]
\State \textbf{Inputs}: Strategy set $\mathcal{K}_{\ell}$ with $|\mathcal{K}_{\ell}|= K$, parameter $\eta$ 

\State \textbf{Initialize} weights: $\mathbf{w}_{\ell,1} = \frac{1}{K}(1,\ldots,1)\in\mathbb{R}^{K}$ 

\State \textbf{for} auction rounds $t = 1$ to $T$, \textbf{do}
\State \textbf{Sample} bidding action $k=(c_{\ell}^{k},d_{\ell}^{k}) \in \mathcal{K}_{\ell}$ randomly, such that $k \sim \mathbf{w}_{\ell,t}$.
\State \textbf{Observe} auction outcomes $\{x_{\ell,t}, \lambda_t, u_{\ell,t}\}$ by solving (\ref{constraints}).

  \State \textbf{Compute} the reward for every action using the aggregate supply curve  $\forall{i} \in \mathcal{K}_{\ell}$. 
  
  \State \textbf{Update} weights: For $\forall i\in \mathcal{K}_{\ell}$,
  \begin{equation*}
    [\mathbf{w}_{\ell,t+1}]_i = \frac{[\mathbf{w}_{\ell,t}]_i \exp(-\eta(1-[\mathbf{u}_{\ell,t}]_i))}{\sum\limits_{j=1}^{K}[\mathbf{w}_{\ell,t}]_j \exp(-\eta (1-[\mathbf{u}_{\ell,t}]_j))}. 
 \end{equation*}
\State \textbf{end for}
\end{algorithmic}
\end{algorithm}
This no-regret Hedge algorithm is standard in online learning with full information feedback \cite{freund1997decision}. The steps are outlined in Algorithm \ref{alg_cap}. In Step $1$, we provide the input data for bidder $\ell$, including her/his set of actions (i.e., price bid options) $\mathcal{K}_{\ell} = \{(c_{\ell}^{1},d_{\ell}^{1}),..., (c_{\ell}^{K},d_{\ell}^{K})\}$, where $ K = \left|\mathcal{K}_{\ell}\right|$ is the number of bidding options of bidder $\ell$, and $\eta$ is the learning rate. 

Step $2$ initializes the uniform weight vector for each bidding option $\mathbf{w}_{\ell,1}\in \mathbb{R}^{K}$. The main iteration procedure of the algorithm is shown in Steps $3$–$7$, which are repeated at each auction round $t$. In Step $4$, given the weights $\mathbf{w}_{\ell,t}$, the algorithm chooses a bidding option $k_{\ell}(t)=(c_{\ell}^{k},d_{\ell}^{k})$ sampled from  a distribution defined by the weights. This sampled action is the one that the bidder submits to the auction in round $t$. The distribution is initialized uniformly and is updated iteratively as will be described, based on the received utilities. The rationale behind the algorithm is to update weights $\mathbf{w}_{\ell,t}$ over rounds, such that bidding options with potentially higher cumulative payoffs get greater weights.

Based on the bids played $ k=(c_{\ell}^{k},d_{\ell}^{k}\in \mathcal{K_{\ell}})$  in Step $4$, in Step $5$, bidder $\ell$ observes the outcomes of the auction for round $t$. These observations include the market-clearing price $\lambda_t$ (public information), the bidder's allocation $x_{\ell,t}$, and the payoff $u_{\ell,t}$ corresponding to the chosen action $k \in \mathcal{K}_{\ell}$ at round $t$. Assuming a posteriori information on the rival bids, in Step $6$, bidder $\ell$ can calculate the payoffs that could have been obtained if alternate bids had been submitted in round $t$. In Step $7$, the algorithm updates the weights of each bidding strategy for bidder $\ell$ based on their performance in round $t$. The update is based on the exponential weight rule, so as to decrease the weight of poor performing actions severely. This expression is then normalized by dividing it by the sum of the updated weights of all actions, ensuring the weight vectors define a probability distribution on the set of available actions of the bidder. The idea is to incrementally shift the weights towards more successful actions (with higher utilities), thereby nudging the bidder towards improved bidding choice over time. With $\eta = \sqrt{\frac{8 \log(K)}{T}}$ the Hedge algorithm is known to obtain the best regret rates in online learning \cite{freund1997decision}. 
 
\section{Benchmark}\label{bench}
In the current literature of day-ahead electricity markets, sequential bidding decisions are mostly addressed through bi-level programming, representing single-leader single-follower or single-leader multi-follower Stackelberg games. As our benchmark, we consider a multi-leader single-follower game. That is, a set of strategic bidders aims to maximize their payoffs, while the market operator seeks to maximize social welfare using an economic dispatch model. In such hierarchical games, achieving a Nash equilibrium between the leader(s) and the follower(s) is generally challenging \cite{ leyffer2010solving, conitzer2011commitment, basilico2017bilevel}. As bi-level programming is widely used in the literature for the bidding strategy problem, we adopt it as a benchmark for the proposed no-regret learning algorithm. In this setting, the assumption is that bidders know the market operator's response before their actions \cite{ruiz2009pool}. 

The general solution method for bi-level programming is to transform the problem into an equivalent single-level optimization problem, which can also be interpreted as a mathematical program with equilibrium constraints (MPEC) \cite{mpecralph}, in a single-leader case. To achieve this, as it is common in the literature \cite{pozo2017basic}, we replace the follower's optimization problem with its Karush–Kuhn–Tucker (KKT) optimality conditions. To accommodate multiple leaders, the MPEC model can then be extended to a multi-leader single (multi) followers problem, known as an EPEC \cite{Ehrenmann}. Solving EPECs is challenging due to their inherent non-convexity, coupled constraints, and the interplay of multiple equilibria within a single problem. In the literature, iterative ways exist to solve this class of problems such as the diagonalization method \cite{diag0,diag}. This iterative approach is what we use, and will be outlined briefly.

\begin{algorithm}[t]
\caption{Diagonalization algorithm to solve the EPEC}
\label{algo_diagonalization}
\begin{algorithmic}[1]
    \State \textbf{Initialize:} $\mathcal{B}_0 = \{c_{\ell}^{k}, d_{\ell}^{k}\quad \forall \ell\};$
     maximum number of iterations $\mathcal{I}$; convergence criterion $\varepsilon$.
    \For{iterations $i = 1$ to $\mathcal{I}$} 
        \For{bidders $\ell = 1$ to $|\mathcal{N}|$}
            \State  Solve  bidder $\ell$'s MIQP given rival strategies.
            \State  Update $\mathcal{B}_{\ell,i}$;
        \EndFor
        \If{$\| \mathcal{B}_{\ell,i} - \mathcal{B}_{\ell,i-1} \| \leq \epsilon \quad\forall \ell \in |\mathcal{N}|$}
            \State Convergence achieved, stop the algorithm.
        \EndIf
        \If{$i = \mathcal{I}$}
            \State Convergence failed, stop the algorithm.
        \EndIf
    \EndFor
\end{algorithmic}
\end{algorithm}

 We consider $|\mathcal{N}|$ strategic bidders who can be leaders and take their decisions first by anticipating the market operator's response. We also consider the market operator who clears the market as a common follower and reacts to the decisions of the upper-level strategic bidders. The objective of each bidder $\ell$ is to maximize her/his own payoff by anticipating how the market operator reacts to her/his decision. On the other hand, the market operator in the lower-level problem minimizes the social cost, i.e., the total cost of meeting demand, accounting for the bid received by $\ell$ and her/his rivals $-\ell$, subject to market constraints. 
The bi-level problem for each bidder $\ell$ is expressed below
\begingroup
\allowdisplaybreaks
\begin{subequations}\label{2}
\begin{align}
\small
\max_{\Xi^{\text{Upper}}}\quad & \lambda x_{\ell} - \big(\frac{1}{2}c_{\ell} x_{\ell}^2 + d_{\ell} x_{\ell}\big) \label{2a}\\
\textrm{s.t.}\quad & c^{k}_{\ell} \in \mathcal{C}_{\ell}, d_{\ell}^{k} \in \mathcal{D}_{\ell} \label{2b} \\
 \min_{\Xi^{\text{Lower}}}\quad  &\frac{1}{2}c_{\ell}^{k} x_{\ell}^2 + d_{\ell}^{k} x_{\ell} + \sum_{-\ell} \big(\frac{1}{2}c_{-\ell}^{k} x_{-\ell}^2 + d_{-\ell}^{k} x_{-\ell}\big) \label{2c}  \\
  \textrm{s.t.} \quad & x_{\ell} + \sum_{-\ell}x_{-\ell} = Q:\quad \lambda\quad \label{2d}  \\
 & 0 \leq x_{\ell} \leq  \Bar{x}_{\ell}:\quad  \underline{\mu}_{\ell}, \bar{\mu}_{\ell}\quad \label{2e} \\
  &  0 \leq x_{-\ell} \leq  \Bar{x}_{-\ell}:\quad  \underline{\mu}_{-\ell}, \bar{\mu}_{-\ell}\quad \forall -\ell, \label{2f}
\end{align}
\end{subequations}
\endgroup
where $\Xi^{\text{Upper}}\!=\!\{c_{\ell}^{k}, d_{\ell}^{k}\}$ is the set of decision variables of the upper-level bidder, $\Xi^{\text{Lower}}\!=\!\{x_{\ell}, x_{-\ell}\}$ is the set of primal variables for the lower-level problem, and $\Xi^{\text{dual}}\!=\!\{\lambda,\bar{\mu}_{\ell}, \underline{\mu}_{\ell}, \bar{\mu}_{-\ell},\underline{\mu}_{-\ell}\}$ represents the corresponding dual variables of the lower-level problem. The first term in the objective function \eqref{2a} of bidder $\ell$ is her/his revenue, i.e., the market-clearing price $\lambda$ multiplied by the production quantity $x_{\ell}$. The second term is the true quadratic cost function for producing $x_{\ell}$. The upper-level constraint \eqref{2b} defines the potential strategic actions of the bidder, such that the bidder can bid her/his cost function with coefficients $c^{k}_{\ell}$ and $d^{k}_{\ell}$, which are not necessarily identical to true coefficients $c_{\ell}$ and $d_{\ell}$. The lower-level objective function \eqref{2c} minimizes the social cost from the perspective of the market operator. By this, we implicitly assume that bidder $\ell$, who solves bi-level program \eqref{2}, perfectly knows the bidding decisions of her/his rivals $-\ell$. 

Notice that from the perspective of a single bidder, an optimizer of the above optimization problem would correspond to her/his \emph{best response} action. Computing the best response action in general is intractable. Thus, this bi-level problem can be reformulated as an MPEC by replacing \eqref{2c}-\eqref{2f} by its KKT conditions. After reformulations proposed by \cite{ruiz2009pool}, the problem is cast as a mixed-integer quadratic program (MIQP). 
The resulting EPEC can be solved using a diagonalization algorithm, an iterative process whose details are given in Algorithm \ref{algo_diagonalization}. Note that the iterations imply that bidders one by one solve (or approximate) their best-response actions. In the case in which they solve the MIQP problem exactly in each iteration, and that the algorithm converges, it follows that the resulting bidding strategy is a Nash equilibrium.

\begin{table*}[t]
\centering
\caption{Bidding options. While \{$c_1, d_1$\} represents the true cost function of each bidder, \{$c_o, d_o$\}, $\forall{o}=\{2,...,10\}$ are her/his available nine options for placing a strategic bid.}
\resizebox{\textwidth}{!}{%
\begin{tabular}{c*{10}{c}*{10}{c}}
\hline
Bidder & $c_1$ & $c_2$ & $c_3$ & $c_4$ & $c_5$ & $c_6$ & $c_7$ & $c_8$ & $c_9$ & $c_{10}$ & $d_1$ & $d_2$ & $d_3$ & $d_4$ & $d_5$ & $d_6$ & $d_7$ & $d_8$ & $d_9$ & $d_{10}$ \\
\hline
$1$ & 0.070 & 0.080 & 0.090 & 0.100 & 0.120 & 0.075 & 0.085 & 0.095 & 0.150 & 0.170 & 9 & 10 & 11.5 & 14 & 13 & 10 & 12 & 13 & 15 & 11 \\
$2$ & 0.020 & 0.050 & 0.060 & 0.150 & 0.250 & 0.025 & 0.150 & 0.900 & 0.130 & 0.310 & 10 & 15 & 12 & 17 & 11 & 12 & 14 & 13 & 16 & 14 \\
$3$ & 0.030 & 0.040 & 0.060 & 0.120 & 0.140 & 0.095 & 0.080 & 0.090 & 0.210 & 0.270 & 12 & 14 & 13 & 16 & 15 & 13 & 12 & 15 & 14 & 12 \\
$4$ & 0.008 & 0.010 & 0.075 & 0.240 & 0.310 & 0.080 & 0.090 & 0.050 & 0.110 & 0.140 & 12 & 14 & 17 & 15 & 17 & 14 & 11 & 17 & 15 & 12 \\
$5$ & 0.010 & 0.090 & 0.100 & 0.210 & 0.970 & 0.020 & 0.130 & 0.075 & 0.190 & 0.095 & 11 & 13 & 11 & 17 & 20 & 12 & 11 & 15 & 17 & 20 \\
\hline
\end{tabular}%
}
\label{tab_1}
\vspace{-3mm}
\end{table*}

\textbf{Remark:} In cases where the number of bidding options for each bidder is small, an alternative approach to solving \eqref{2} would be an enumeration process. For instance, consider a scenario with three bidding options for every bidder. In every iteration, with fixed rival bids, each bidder $\ell$ solves three quadratic programs (QPs) \eqref{constraints} (which could be solved in parallel)—one for each bidding option, calculates her/his payoff and chooses the best response accordingly. This approach requires each bidder to solve one QP per bid option in every iteration, thus bypassing the need to solve a single MIQP in every iteration of Algorithm \ref{algo_diagonalization}.

\vspace{5mm}
\section{Numerical Study}\label{case}

We consider five bidders, namely Bidders $1$ to $5$, participating in an hour-ahead electricity market. Since no inter-temporal constraint is enforced in the market-clearing optimization problem \eqref{constraints}, the auction for different hours can be conducted separately and independently. This enables us to focus on the bidding strategy for a specific hour and base our analysis on the hour-ahead market \cite{chen2019learning}. We fix the total load $Q$ to be $1148.4$ MW, which is the total demand in the Danish bidding zone DK$1$ reported by Nord Pool for hour $9$:$00$-$10$:$00$ on August $27$, $2023$. Table \ref{tab_1} presents the action sets. For each bidder, we consider ten options for coefficients \{$c$, $d$\}, where \{$c_1, d_1$\} refers to her/his true quadratic cost function. The remaining nine bidding options \{$c_2, d_2$\} to \{$c_{10}, d_{10}$\} let the bidder place a strategic price bid. These nine options are generated by gradually increasing the quadratic and linear terms of the cost function. The capacity of bidders is set identically to $700$ MW. All source codes are provided in \cite{Aregagit}. 

\vspace{2mm}
\subsection{Cases Based on Bidding Approaches}
\vspace{-1mm}

We first start by defining two \textit{benchmark} cases and then introduce six \textit{learning-oriented} cases, resulting in eight cases. The two benchmark cases are as follows:
\begin{itemize} 
\item Case $a$: \textbf{Best Response}: All Bidders $1$ to $5$ have perfect information of rival bids and use this to compute their best response. For this case, we use the diagonalization approach in Algorithm \ref{algo_diagonalization}.
\item Case $b$: \textbf{Trustful}: All Bidders $1$ to $5$ submit their true generation cost \{$c_1, d_1$\}. This results in a perfectly competitive market, for which we solve the market-clearing optimization problem \eqref{constraints} given true costs.
\end{itemize}

We now define six learning-oriented cases. In all these six cases, Bidder $5$  employs either the proposed Hedge algorithm \ref{alg_cap} (cases $c$, $e$, and $g$) to learn her/his strategic price bids over auction rounds or bids randomly (cases $d$, $f$ and $h$). Conversely, rival bidders (Bidders $1$-$4$) utilize their true cost bids (cases $c$ and $d$), learn using the Hedge algorithm (cases $e$ and $f$), and bid randomly (cases $g$ and $h$):
\begin{itemize} 
\item Case $c$: \textbf{Trustful vs Hedge}: We assume Bidders $1$-$4$ are naive and provide their true cost bids in every round, while Bidder $5$ learns by employing the Hedge algorithm. That means, only Bidder 5 is allowed to choose her/his bids from the ten options provided in Table \ref{tab_1} and learns how to bid strategically based on the Hedge algorithm. 
\item Case $d$: \textbf{Trustful vs Random}: In this case, instead of using the Hedge algorithm, Bidder 5 bids randomly, while rival bidders consistently submit their true costs. 
\item $\rm{Case }$ $e$: \textbf{Hedge vs Hedge}: Both Bidder 5 and her/his rival bidders (Bidders $1$-$4$) utilize the Hedge algorithm to simultaneously learn their strategic bids.
\item Case $f$: \textbf{Hedge vs Random}: Bidders $1$-$4$ employ the Hedge algorithm, while Bidder $5$  bids randomly.
\item Case $g$: \textbf{Random vs Hedge}: Bidders $1$-$4$  choose their bids uniformly at random in each round of the auction while Bidder $5$  exploits the proposed Hedge algorithm. 
\item Case $h$: \textbf{Random vs Random}: All bidders sample bids uniformly at random in each round of the auction from bidding options in Table \ref{tab_1}.
\end{itemize}

\begin{table*}
\centering
\caption{Social cost [\euro] for various cases in increasing order.}
\resizebox{\textwidth}{!}{%
\begin{tabular}{|l|l|l|l|l|l|l|l|l|}
\hline
Cases & Trustful & Trustful vs Hedge & Trustful vs Random & Best Response & Hedge vs Hedge & Hedge vs Random & Random vs Hedge & Random vs Random \\ \hline
Social cost [\euro] & $19,419$ & $21,085$ & $21,332$ & $24,408$ & $26,162$ & $28,799$ & $29,292$ & $35,763$ \\ \hline
\end{tabular}%
}
\label{Tab_2}
\end{table*}

For all the six learning cases, we use the optimization problem (\ref{constraints}) to simulate the auction environment and determine the bidders' bidding strategies over  $T= 200$ rounds of the auction. Consequently, we run these $200$ auction rounds $15$ times to calculate the average outcomes across different runs. Hereafter, solid lines in our figures show the average values over $15$ runs, whereas shaded areas cover the mean value plus and minus the standard deviation. Unlike the six learning-oriented cases ($c,d,e,f,g,h$), the benchmarks (cases $a,b$) are not built upon data gathered from learning algorithms, since bidders either bid their true costs or submit their a priori computed best response bids. 

\vspace{2mm}
\subsection{Market Outcomes: System-level Results}
\vspace{-1mm}
 
For all eight cases, we report the social cost, denoting the social welfare, and the market-clearing price. Recall that the social cost is the optimal value of objective function \eqref{constraints1}, and the market-clearing price is the optimal value of the dual variable associated with \eqref{constraints2}. We note that the EPEC problem solved by the diagonalization Algorithm \ref{algo_diagonalization} converged after four iterations with $\epsilon = 0.0004$. Furthermore, through enumeration, we verified that the converged bid profiles were a Nash equilibrium of the game.

Table \ref{Tab_2} provides the social cost of the eight cases, arranged in ascending order. 
We discuss the results of this table together with those plotted in Figure \ref{result}, illustrating the evolution of social cost (left panel) and that of the market-clearing price (right panel) over $200$ auction rounds. The social cost results in Table \ref{Tab_2} correspond to those in the last round of the upper plot of Figure \ref{result}. As expected, the social cost is lower in the $\rm{Trustful}$ than the cases in which Bidder $5$ is acting strategically through the Hedge algorithm or randomly. Compared to the $\rm{Best}$ $\rm{Response}$ where bidders are using their Nash equilibrium strategy, the social cost increases when one or all bidders adopt a learning approach. Furthermore, with the increase of randomness (zero, one, and all bidders applying $\rm{Random}$ strategy), the social cost increases. 

Given that implementing the Hedge algorithm and computing the best response strategies require progressively more information than playing a random strategy, the results above quantify the value of information to the bidders, and suggest that having more information increases market efficiency. %Furthermore, given that all players implementing the Hedge algorithm results in a higher social welfare than that corresponding to $\rm{Best} \rm{Response}$, we conclude the coarse-correlated equilibrium the algorithm is converging to less efficient than the one corresponding to the Nash equilibrium learned.

\begin{figure*}[t] 
  \centering
        \includegraphics[width=3.4in]{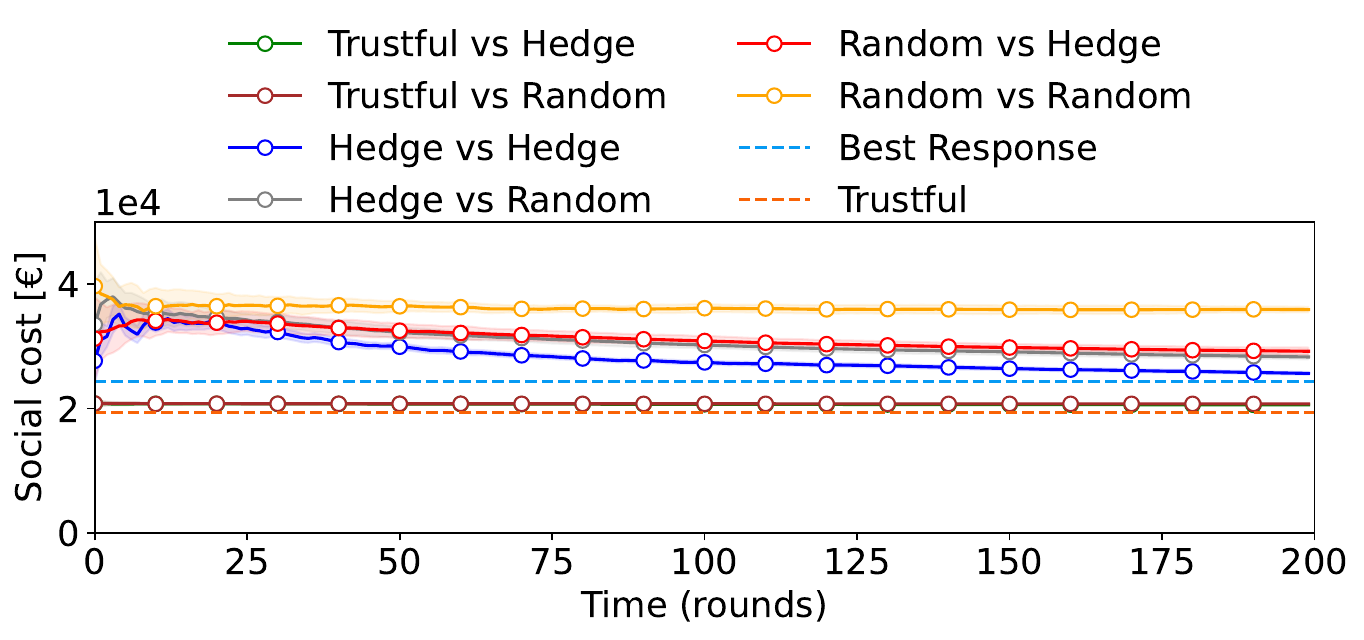} 
        \includegraphics[width=3.4in]{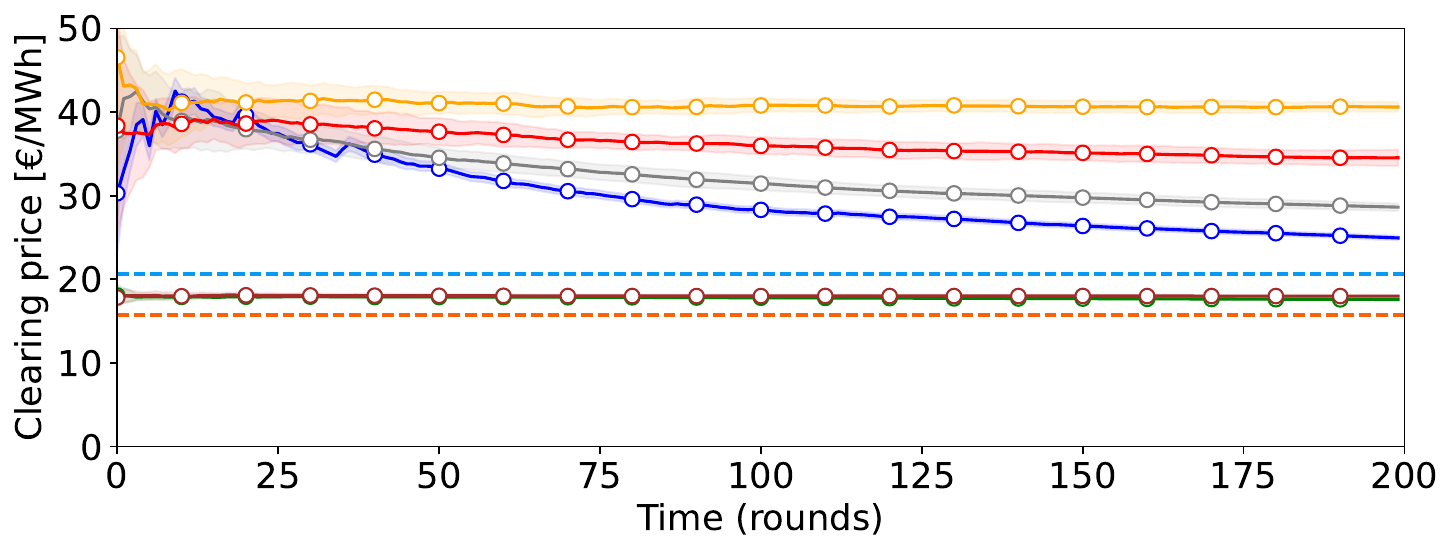} 
\caption{Evolution of social cost (left panel) and market-clearing price (right panel) over $200$ auction rounds.}
\label{result}
\vspace{-2mm}
\end{figure*}

\begin{figure*}[t] 
  \centering
        \includegraphics[width=3.4in]{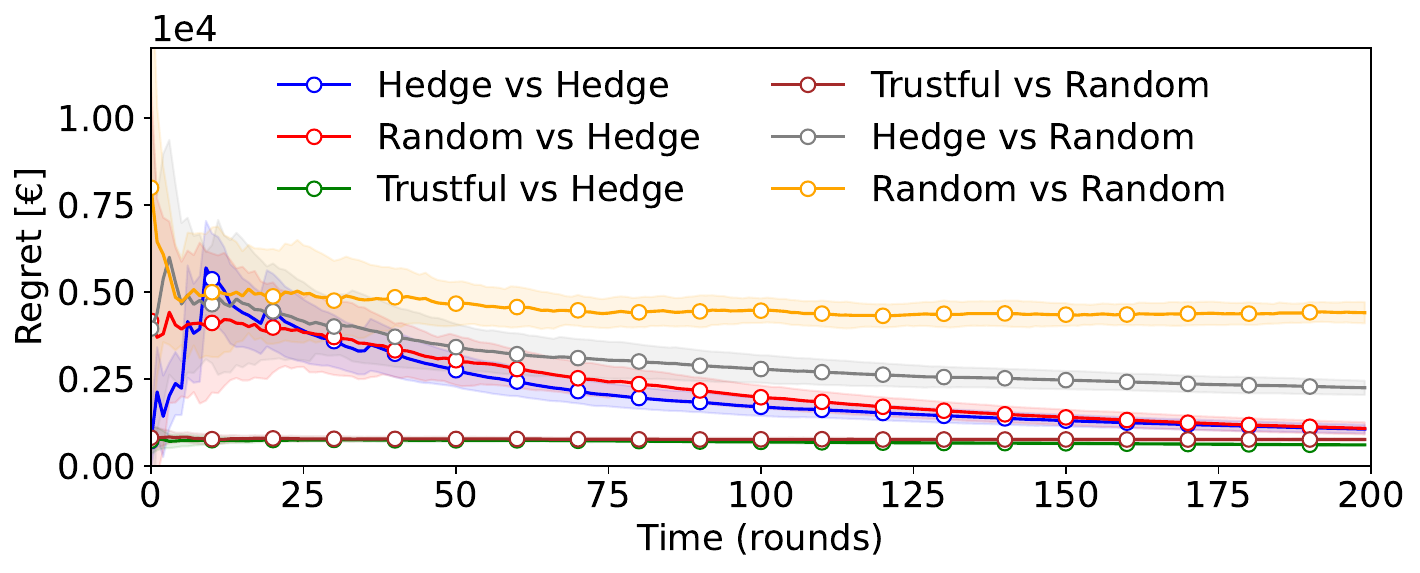}  
        \includegraphics[width=3.4in]{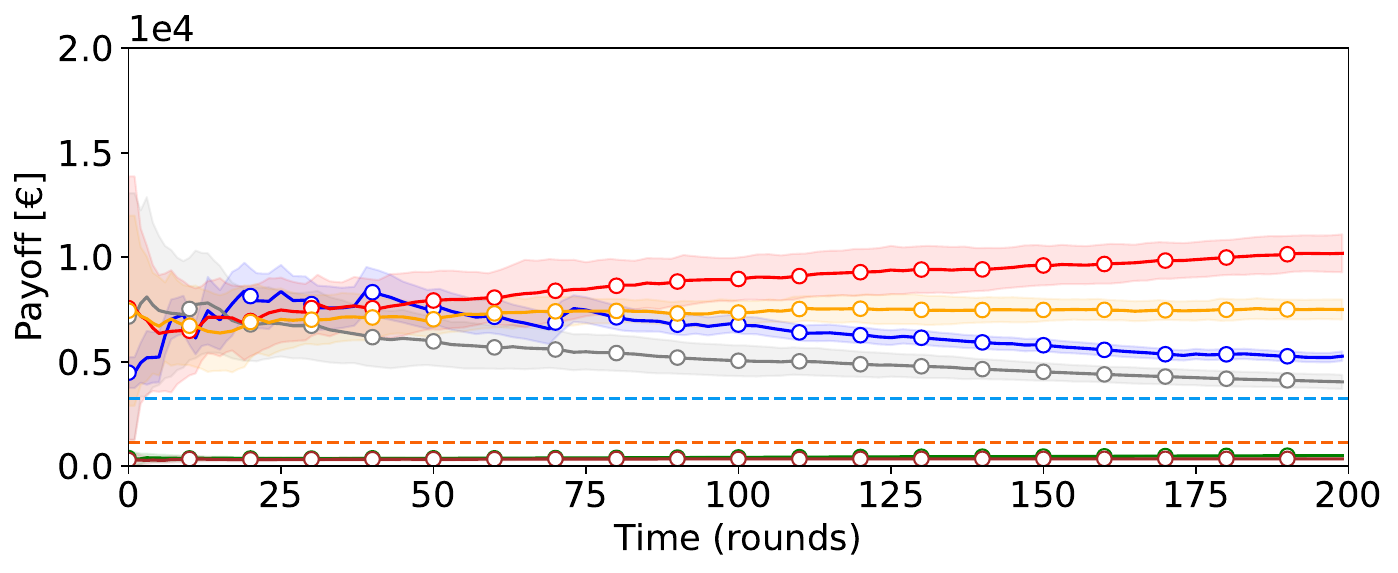} 
  \caption{Evolution of average regret (left panel) and average payoff (right panel) for Bidder $5$ over $200$ auction rounds.}
  \label{result2}
  \vspace{-3mm}
\end{figure*}

\vspace{3mm}
\subsection{Market Outcomes: Individual Results}
\vspace{-1mm}
Figure \ref{result2}, left panel, illustrates the average regret of Bidder $5$ in the six learning-oriented cases. As expected, when rivals stay with a given approach (Hedge algorithm, or randomization), Bidder $5$'s regret decreases when employing the Hedge algorithm in comparison to playing randomly. 

Figure \ref{result2}, right panel, illustrates the payoff of Bidder $5$. From this figure, we make three observations below. 

First, if rivals bid randomly, Bidder $5$ earns a higher payoff, regardless of her/his bidding strategy, compared to other cases where rivals do not submit randomly. When Bidder $5$ employs the Hedge algorithm while rival bidders act randomly (the red curve), her/his payoffs are higher than if she/he were to also use a random strategy (yellow curve).
     
Second, if rivals employ the Hedge algorithm, it is slightly better for Bidder $5$ to use the Hedge algorithm too (blue curve) compared to bidding randomly (gray curve). While this is expected, it is interesting to observe that in these two cases, the average regret for Bidder $5$  is decreasing despite the fact that her/his payoffs are decreasing. While the decreasing trend of regret by the learning algorithm is expected, in this auction setting higher regret does not mean a lower payoff (see orange line) and lower regret does not mean a higher payoff (see blue). Hence, this suggests that regret may not be the best benchmark for the problem at hand and it is more important to directly measure the bidders' utilities and social costs.

Third, if rival bidders submit bids based on their $\rm{Trustful}$ bids, Bidder $5$ earns a higher payoff by bidding her/his true costs (red dashed line) compared to when Bidder $5$ employs the Hedge algorithm (green) or bids randomly (brown). The results show that Bidder $5$’s payoff, when using the Hedge algorithm (green), while her/his rivals consistently use their $\rm{Trustful}$ bidding, \textit{slowly} converges to all $\rm{Trustful}$ bidding (red dashed line). This particular observation is due to the fact that by enumeration, we could verify that Bidder $5$'s best response to others' playing their true costs is to play her/his true cost.

\vspace{3mm}
\section{Conclusion}\label{conc}
This paper presented a no-regret learning approach to analyze the extent of market power that bidders can exert by learning their strategic bids in the day-ahead electricity market. To this end, the Hedge algorithm, which is known to obtain the best regret rates in online learning, is utilized. The proposed learning approach shows promising results by relying solely on the posterior information published by the market operator, without making any assumptions about rival bids a priori in the bidding decision making process. 

From the system perspective, we observed that electricity markets might be exposed to greater market power compared to the case in which bidders provide their best responses, as classical game theoretic approaches would result in, but use a learning approach. We further observed the less information bidders have to choose their actions, the higher the social welfare will be.

From a bidder’s perspective, we illustrate the decreasing regret of a single bidder over auction rounds. However, bidders' payoffs by using the Hedge algorithm highly depend on rivals' strategies. In particular, unless other bidders are playing a fixed strategy, a given bidder has a higher payoff if she/he uses the proposed online learning approach. However, if other bidders play a fixed strategy over auction rounds (submitting either their true costs or their best response strategies), then there is clearly a single fixed strategy that outperforms no-regret learning and in this case, the no-regret algorithm converged to the corresponding strategy. Furthermore, from our case study, we observed that a bidder earns a higher payoff when she/he is the only one who employs a learning approach, and rival bidders submit their bids randomly, compared to the cases where everyone is learning simultaneously. Lastly, we observed that a decrease in regret in one player may not imply an increase in the payoffs of the player and vice versa. This motivates use of other metrics such as policy regret to address learning in repeated electricity market auctions.

Our formulations and case studies serve as a preliminary evaluation of the potential of the online learning approach in electricity markets. There are several directions for future research. First, a large number of simulations with different bidders and market constraints are needed to generalize the observations made in our case study to realistic market settings. Second, in order to simulate the full information feedback considered in our case study, the aggregate supply curves published a posteriori need to be utilized. Third, in a realistic market setting, contextual information such as weather forecasts could complement a bidder's information and thus, it is important to leverage this information in online learning to improve a bidder's performance. Fourth, the auction environment can have multiple trading stages such as spot and balancing markets.

\vspace{3mm}
\section*{Acknowledgments}
Arega Getaneh Abate has received funding from the European Union’s Horizon $2020$ research and innovation program under the Marie Skłodowska-Curie grant agreement No $899987$. Maryam Kamgarpour gratefully acknowledges the project funding under ``UrbanTwin" with the financial support of the ETH-Domain Joint Initiative program in the Strategic Area Energy, Climate and Sustainable Environment. 
\vspace{4mm}

\bibliographystyle{IEEEtran}
\bibliography{reference}

\end{document}